**Phase Transitions and Topological Protection in Anyonic-PT-Symmetric Lattices**

*Ruiying Zhang[1], Ziteng Wang[1], Daohong Song[1,2], Liqin Tang[1,2*], Konstantinos G. Makris[3,4*], and Zhigang Chen[1,2*]*

Ruiying Zhang and Ziteng Wang contributed equally to this work.

Ruiying Zhang, Ziteng Wang, Daohong Song, Liqin Tang, Zhigang Chen
The MOE Key Laboratory of Weak-Light Nonlinear Photonics, TEDA Applied Physics Institute and School of Physics, Nankai University, Tianjin 300457, China
*E-mail: tanya@nankai.edu.cn; makris@physics.uoc.gr; zgchen@nankai.edu.cn

Liqin Tang, Daohong Song, Zhigang Chen
Collaborative Innovation Center of Extreme Optics, Shanxi University, Taiyuan, Shanxi 030006, China

Konstantinos G. Makris
ICTP, Department of Physics, University of Crete, 71003 Heraklion, Greece
Institute of Electronic Structure and Laser (IESL), FORTH, 71110, Heraklion, Greece

Funding: National Key R&D Program of China (2022YFA1404800); National Natural Science Foundation of China (12134006, W2541003, 12374309, 12274242 and 124B2078); Natural Science Foundation of Tianjin (21JCJQJC00050); the 111 Project (B23045) in China; Institute of Electronic Structure and Laser, FORTH and the University of Crete by the European Research Council (ERC-Consolidator) under the grant agreement No. 101045135 (Beyond_Anderson).

**Abstract:** Parity-time (PT) symmetry and anti-PT symmetry have attracted extensive interest for their non-Hermitian spectral properties, particularly the emergence of purely real and imaginary eigenvalues in their symmetry-unbroken regime, respectively. Recently, these two scenarios have been unified under a more general framework known as anyonic-PT symmetry, yet its physical implications in waveguide platforms and corresponding topological features in extended lattice systems remain largely unexplored. Here, the phase transitions and topological protection in anyonic-PT-symmetric systems are systematically investigated in waveguide lattices. In the symmetry-unbroken regime, the arguments of all bulk eigenvalues are constrained to two discrete values separated by $\pi$, leading to distinctive oscillatory propagation dynamics accompanied by controlled amplification or dissipation. In the case of one-dimensional lattice, the energy bands exhibit a gap closing and reopening during phase transition. Moreover, in the symmetry-unbroken regime, the topological edge states emerge within the bulk gap and are protected by a generalized pseudo-anyonic-Hermiticity (PAH) symmetry. Our results establish anyonic-PT symmetry as a new tunable degree of freedom for non-Hermitian waveguide systems, where the eigenvalue argument provides a natural quantity for information encoding. This work broadens the conceptual foundation of topological protection under generalized non-Hermitian symmetries.

## 1. Introduction

Topology has emerged as a unifying framework in modern physics, offering profound insights into the behavior and control of diverse physical systems.[1–6] A cornerstone of this framework is the bulk-boundary correspondence, which establishes a direct relationship between bulk topological invariants and the emergence of robust boundary states. These topological boundary states exhibit remarkable resilience to perturbations that preserve specific symmetries, thereby manifesting topological protection. In recent years, substantial progress has been made in topological physics, with studies extending beyond Hermitian systems to the non-Hermitian regime.[7–18] The exploration of topology in the context of parity-time (PT) symmetry[19–21] and non-Hermitian photonics,[22–30] where gain or loss naturally arise in non-conservative systems, offers a promising direction for both deepening theoretical understanding and enabling practical control in complex physical systems.

Non-Hermitian systems, typically exhibiting complex eigenvalues and non-orthogonal eigenstates,[19–21] support a range of phenomena with no Hermitian counterparts, including PT symmetry phase transition,[22–31] exceptional points (EPs),[32–41] and non-Hermitian skin effect (NHSE).[42–49] On the other hand, symmetry plays a pivotal role in physics, representing invariance and conservation laws under specific transformations. In non-Hermitian systems, PT and anti-PT symmetry[50–53] have attracted growing attention for their ability to yield purely real or imaginary eigenvalues in the symmetry-unbroken regime. A phase transition occurs precisely when eigenstates no longer respect the underlying symmetry. The critical point corresponds to an EP, which exhibits a unique type of non-Hermitian degeneracy.[54–59] Quite recently, a new concept called anyonic-PT symmetry has been introduced,[60] which unifies PT and anti-PT symmetries through a tunable phase factor. So far, anyonic-PT symmetry has been applied only to a limited number of studies such as in complex-coupled lasers,[61] entropy dynamics in quantum systems,[62] and tunable EPs in coupled resonators systems.[63] However, these studies mainly concentrate on systems comprising only a few coupled elements and emphasize the observable phenomena. The phase transition induced by anyonic-PT symmetry itself has not been systematically discussed. To the best of our knowledge, such symmetry has not yet been implemented in extended waveguide lattices, and the associated topological phenomena need further investigation.

In this work, we investigate the phase transition under anyonic-PT symmetry in non-Hermitian waveguide periodic systems. Through theoretical analysis and a simple two-site model, we reveal the properties of the symmetry-unbroken regime, where the eigenvalue arguments are restricted to two fixed values differing by $\pi$. The optical power exhibits oscillation with overall amplification or dissipation. Then, we extend the system to a one-dimensional (1D) waveguide lattice and find that the parameter space can be divided into three distinct phases: *unbroken, partially broken, and fully broken*. During the phase transition, energy bands undergo a gap closing and reopening. When the coupling is nontrivial, the system supports two topological edge states. In the symmetry-unbroken regime, the edge states reside in the gap and are protected by a new symmetry, termed pseudo-anyonic-Hermiticity (PAH) symmetry. This symmetry protects both the chiral character and the energy phase of the edge states.

**2. Anyonic Parity-time Symmetry**

Anyonic-PT symmetry is defined by the commutation relation:

$$PTH = \exp(-2i\Gamma)\, HPT, \tag{1}$$

where $P$ and $T$ denote parity and time-reversal operators, respectively. $\Gamma$ is a real parameter in the range $[0, \pi)$ and controls the symmetry of the system. This means that, when $\Gamma = 0$, the relation reduces to the PT-symmetric case, while for $\Gamma = \frac{\pi}{2}$, it corresponds to the anti-PT-symmetric case.

More specifically, in a time-independent (or homogeneous along propagation distance) system, the eigenvalues $E$ and eigenstates $\psi$ are governed by the time-independent Schrödinger equation $H\psi = E\psi$. For any given Hamiltonian H, if it satisfies the anyonic-PT symmetry condition described by Equation (1), the eigenvalues exhibit the following property:

$$E = \exp(2i\Gamma)E^*. \tag{2}$$

The asterisk denotes complex conjugation and represents time reversal, which is the result of the action of time-reversal operator. We can express the eigenvalues in polar form as $E = A\exp(i\phi)$, where $\phi$ is the argument of a complex number. Substituting this polar form into Equation (2) yields a simple yet clear condition on $\phi$:

$$\phi = \Gamma + n\pi, \tag{3}$$

where n is an integer. Equation (3) describes the key feature of the system in anyonic-PT symmetry-unbroken regime. That is, the arguments of the eigenvalues can only take on two fixed values: $\Gamma$ and $\Gamma+\pi$. This generalization not only unifies underlying mathematical structures but also unveils unconventional features across diverse physical systems.

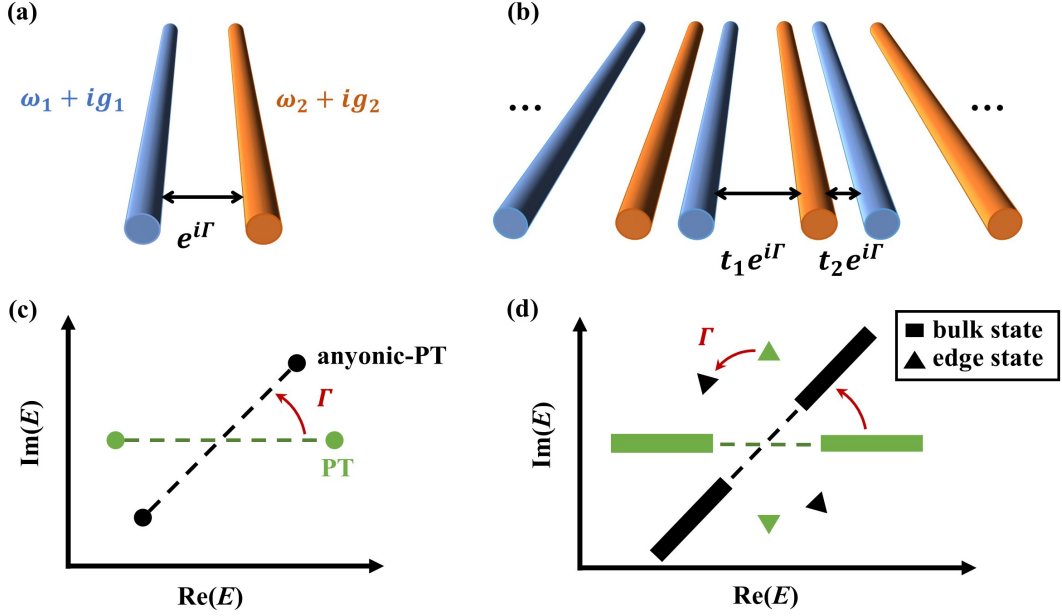

**Figure 1. The Anyonic-PT-symmetric models and their energy characteristics.** (a) Schematic illustration of two-site model under anyonic-PT symmetry. Each waveguide has a complex onsite potential $\omega_i + ig_i$. The coupling between two waveguides is $\exp(i\Gamma)$. (b) A one-dimensional lattice composed of unit cells identical to the model shown in (a), with modulated intracell $t_1\exp(i\Gamma)$ and intercell $t_2\exp(i\Gamma)$ couplings. (c-d) Eigenvalue comparison of the PT-symmetric case (green) and anyonic-PT-symmetric case (black) in the symmetry-unbroken regime. $\Gamma$ is a controllable parameter for anyonic-PT symmetry, which manifests as the rotation angle of the eigenvalues relative to the PT-symmetric axis.

## 3. Phase Transition and Dynamics in a Two-site Model

In this section, we present a simple case of a two-site waveguide model (**Figure** 1a) to investigate the anyonic-PT symmetry phase transition. Here $\omega_i + ig_i$ represents the onsite potential, in which the real part represents the refractive index and the imaginary part corresponds to gain or loss. The Hamiltonian for such a two-site system, described by tight-binding model, can be written as:

$$H = \begin{pmatrix} \omega_1 + ig_1 & \exp(i\Gamma) \\ \exp(i\Gamma) & \omega_2 + ig_2 \end{pmatrix}. \tag{4}$$

If this matrix satisfies the anyonic-PT symmetry, then, by solving Equation (1), the diagonal parameters need to adhere to the following conditions:

$$\omega_2 = \omega_1 \cos 2\Gamma + g_1 \sin 2\Gamma,$$
$$g_2 = \omega_1 \sin 2\Gamma - g_1 \cos 2\Gamma. \tag{5}$$

When $\Gamma = 0$, this model returns to the well-known PT-symmetric case, where $\omega_1 = \omega_2$ and $g_1 = -g_2$. In this case, the two eigenvalues of the system have identical imaginary parts. Furthermore, a non-zero $\Gamma$ will rotate all parameters by an angle $\Gamma$ on the coordinate axis, as Figure 1c shows. This represents the eigenvalue characteristics in the anyonic-PT symmetry-unbroken regime.

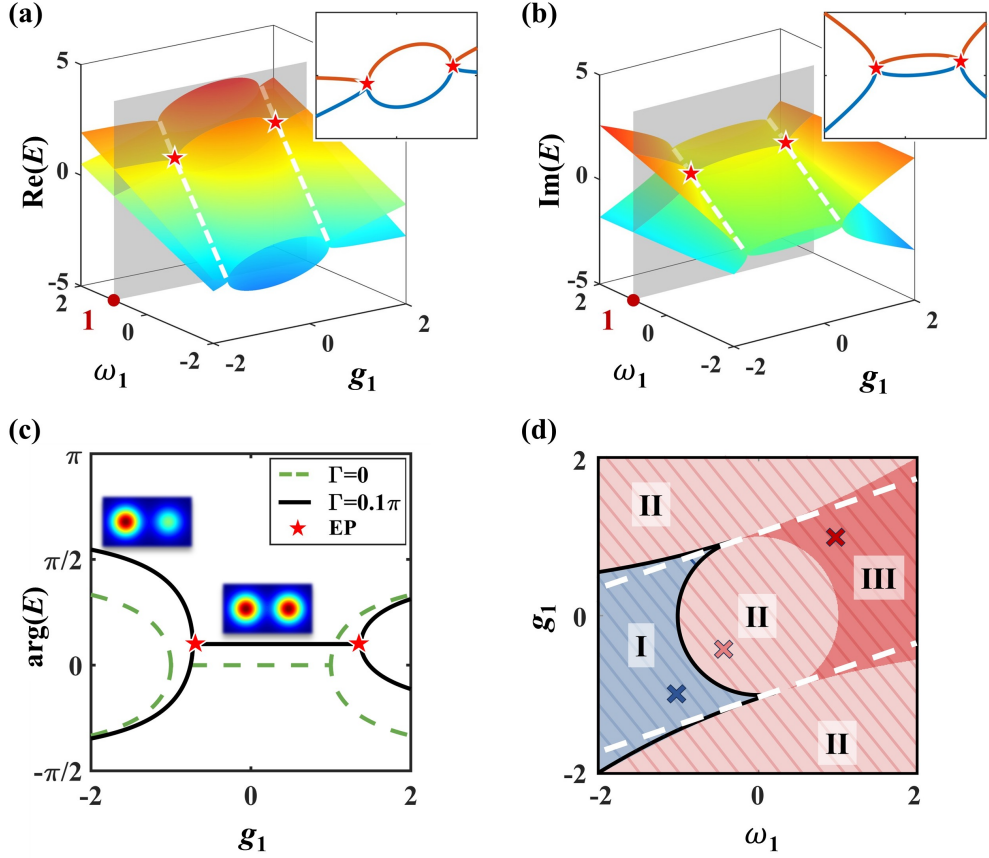

**Figure 2. Eigenvalues and propagation behavior of two-site model.** (a-b) Real part (a) and imaginary part (b) of eigenvalues in the parameter space spanned by $\omega_1$ and $g_1$ for $\Gamma = 0.1\pi$. White dashed lines depict the phase transition lines. The gray plane denotes the truncation surface at $\omega_1 = 1$ section, and its intersection with the eigenvalue is highlighted in the inset. The EPs are marked as red pentagrams. (c) Eigenvalue argument curves under different values of $\Gamma$ at $\omega_1 = 1$. Two EPs occur at $g_1$ approximately equal to -0.7 and 1.3. Insets show the eigenstate amplitude profiles in symmetry-unbroken and broken regimes. (d) Dynamical phase diagram of the anyonic-PT-symmetric system in parameter space. The blue and red regions, which are separated by the black curves, represent dissipation and amplification, respectively. The colored crosses correspond to the parameters chosen in three different regions. Region I: immediate dissipation; Region II: delayed amplification; Region III: immediate amplification.

Next, we explore the characteristics of the eigenvalues within the parameter space defined by $\omega_1$ and $g_1$. Due to the relationship in Equation (5), $\omega_2$ and $g_2$ can be expressed in terms of $\omega_1$ and $g_1$. As shown in **Figure** 2a,b, under a fixed value of $\Gamma = 0.1\pi$, a notable distinction from conventional PT-symmetric cases is that both the real and imaginary parts of eigenvalues exhibit a hollow shape. The two energies intersect along the white dashed lines, which indicate the locations of the anyonic-PT-symmetric phase transition. Along these lines, both the eigenvalues and eigenstates coalesce (see Supporting Information Equation (S1-S3)), forming exceptional lines, where every point corresponds to an EP. The region bounded by the two dashed lines indicates the anyonic-PT symmetry-unbroken regime, whereas the area outside corresponds to the broken phase. In contrast to conventional PT or anti-PT systems—where the phase transition is typically controlled along a single parameter axis—the anyonic-PT phase transition here can be independently tuned along either $\omega_1$ and $g_1$ parameter dimension. For example, by fixing $\omega_1 = 1$, we can see how the eigenvalues change with $g_1$ varying in insets of Figure 2a,b. EPs are marked by red pentagrams. Except at EPs, the real and imaginary parts of the two eigenvalues are non-degenerate, which not only enhances the distinguishability of the corresponding eigenstates, but also enables more flexible and independent control over individual modes within the system.

To intuitively illustrate the phase transition, we present the values of the eigenvalue argument in Figure 2c, comparing anyonic-PT symmetry with conventional PT symmetry. The curves show the evolution of arguments as a function of $g_1$ at $\omega_1 = 1$ for two values of $\Gamma$. The degenerate region of two energies corresponds to the symmetry-unbroken phase. For $\Gamma = 0.1\pi$ (black lines), phase transitions occur at approximately $g_1 \approx -0.7$ and $g_1 \approx 1.3$. The specific parameter values of EPs can be obtained from Equation (S3) in the Supporting Information. Compared with the PT-symmetric case, the EPs are shifted due to the overall rotation of the eigenvalues. Insets depict the corresponding eigenstate magnitude distributions: uniform magnitude across both sites in the symmetry-unbroken regime (middle inset), and asymmetric magnitude in the symmetry-broken regime (left inset).

Although the eigenstates resemble those in conventional PT-symmetric systems, the distinct eigenvalue structure of the anyonic-PT-symmetric system leads to qualitatively different wave propagation dynamics. Figure 2d presents the corresponding dynamical phase diagram in parameter space. This diagram allows us to determine the propagation characteristics under a single-site excitation at an arbitrary site. Here, we focus only on the dynamics in the symmetry-unbroken regime, corresponding to the region between the white dashed lines. The power shows oscillation[23] with dissipation or amplification. The blue region represents dissipation, while the red region denotes amplification. Based on the imaginary parts of the eigenvalues, parameter space can be further divided into three distinct regions, denoted as I, II, and III, each exhibiting different dynamical behaviors. The blue area corresponds to Region I, where the imaginary parts of both eigenvalues are negative, leading to dissipation during propagation. In Region II, one eigenvalue has a positive imaginary part and the other is negative. Therefore, when the initial state is an arbitrary superposition of two eigenstates, the decaying mode gradually vanishes during propagation, while the amplifying mode becomes dominant. During propagation, the overall power may sometimes exhibit an initial dissipation followed by subsequent amplification. In Region III, both imaginary parts are positive, resulting in immediate amplification. We present the power dynamics in different regions in the Supporting Information, corresponding to the three colored crosses in Figure 2d.

## 4. Topological Properties in Dimer Lattices

Now we move on to the anyonic lattices. In particular, we examine a dimer lattice (**Figure** 3a) that satisfies anyonic-PT symmetry. This model can be seen as an extension of a 1D non-Hermitian Su–Schriefer–Heeger (SSH) model.[64–67] By investigating this lattice, we can determine the band structure in momentum space as well as the boundary phenomena in real space.

The Hamiltonian in momentum space can be written in the following form:

$$H_k = \begin{pmatrix} \omega_1 + ig_1 & t_1\exp(i\Gamma) + t_2\exp(i\Gamma)\exp(-ik) \\ t_1\exp(i\Gamma) + t_2\exp(i\Gamma)\exp(ik) & \omega_2 + ig_2 \end{pmatrix}, \quad (6)$$

where $k$ is the wavenumber in the momentum space. The intracell and intercell couplings are $t_1\exp(i\Gamma)$ and $t_2\exp(i\Gamma)$, respectively. Here $t_1$ and $t_2$ are real. Each unit cell, highlighted with red dashed lines in Figure 3a, contains two sites with different onsite potentials. Similar to the two-site model, the onsite potentials are required to obey Equation (5).

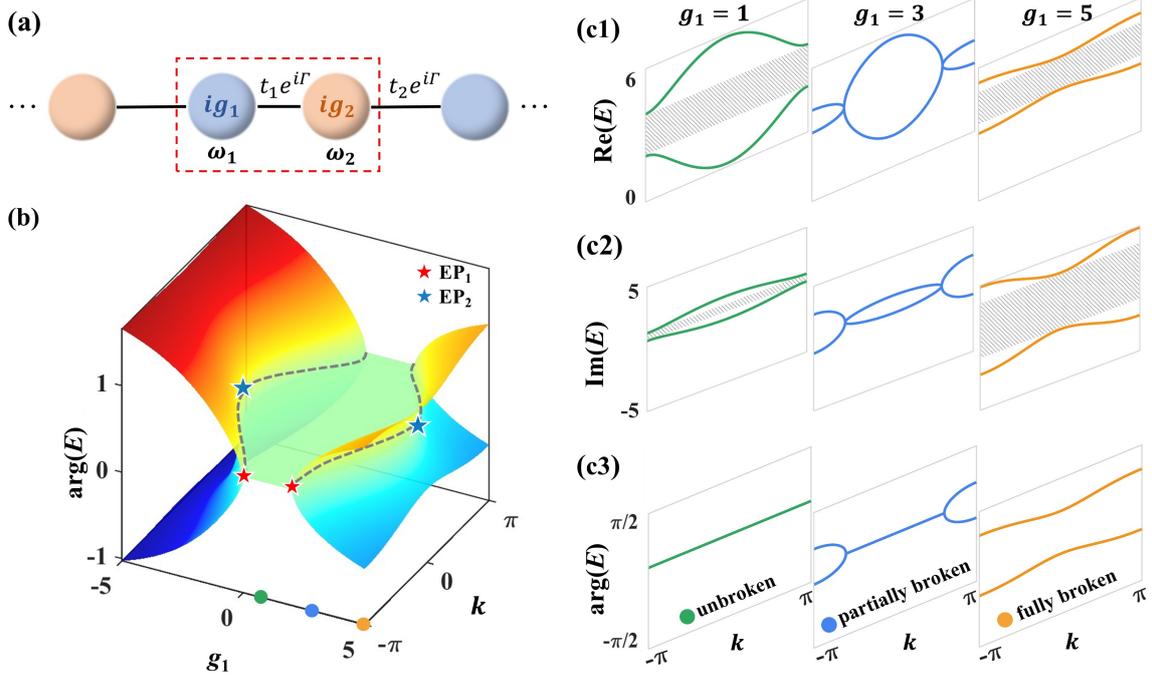

**Figure 3. Infinite dimer waveguide lattice and band structures.** (a) Schematic of an infinite lattice. The unit cell (red dashed line) features onsite potentials $\omega_i + ig_i$, with modulated couplings $t_1\exp(i\Gamma)$ (intracell) and $t_2\exp(i\Gamma)$ (intercell). (b) Eigenvalue arguments in the synthetic $(g_1, k)$ parameter space. EPs, indicated by red and blue pentagrams, are located at the center and boundary of the Brillouin zone (BZ). Three colored points along the $g_1$ axis denote parameter values for distinct phases. (c) Band structures of real part (c1) imaginary part (c2) and argument (c3) as a function of $k$ for different $g_1$ values. The color of each subfigure corresponds to the parameter point marked with the same color in (b). The energy gaps are highlighted by the gray shaded regions. For all calculations here, we use $\Gamma = 0.1\pi$, $\omega_1 = 3$, $t_1 = 1$ and $t_2 = 2t_1$.

Different from the two-site model, the phase transition condition of the dimer lattices depends on both the parameters and Bloch wavenumber (see Supporting Information Equation (S4)). In this system, three distinct phases can be identified: symmetry-unbroken, partially broken, and fully broken, separated by EPs. In Figure 3b, different EPs are marked by red (EP1) and blue (EP2) pentagrams. Here, we still focus on the phase transition as $g_1$ varies and choose $\omega_1$ to be 3, $\Gamma = 0.1\pi$, $t_1 = 1$ and $t_2 = 2t_1$. From Figure 3b, one can find a green flat region in the middle, where the energy argument is uniform. In the region between the two red pentagrams, the energy argument value is constant for all $k$. We refer to this region as the anyonic-PT symmetry-unbroken regime. However, in the region bounded by the red and blue pentagrams, we observe that a portion of the eigenvalue argument remains pinned at a fixed value near the center of the Brillouin zone (BZ), while it begins to vary near the BZ boundary. This indicates that symmetry breaking initiates at the BZ boundary. We refer to this region as the symmetry partially broken regime. Finally, when the selected region is outside the blue pentagrams, the energy argument varies across the entire BZ, corresponding to the symmetry fully broken regime.

Beyond the unique eigenvalue argument behavior, the energy bands undergo a gap closing followed by a reopening during the phase transition. In Figure 3c, we show the band structure for the real part (c1), imaginary part (c2) and argument (c3) of the eigenvalues. The colors of the energy bands correspond to the parameter values indicated by the colored dots in Figure 3b. It can be observed that a gap exists in both the symmetry-unbroken and fully broken phases. This serves as a signature of the phase transition and provides the condition for the edge states to reside in the band gap, thereby enabling topological protection.

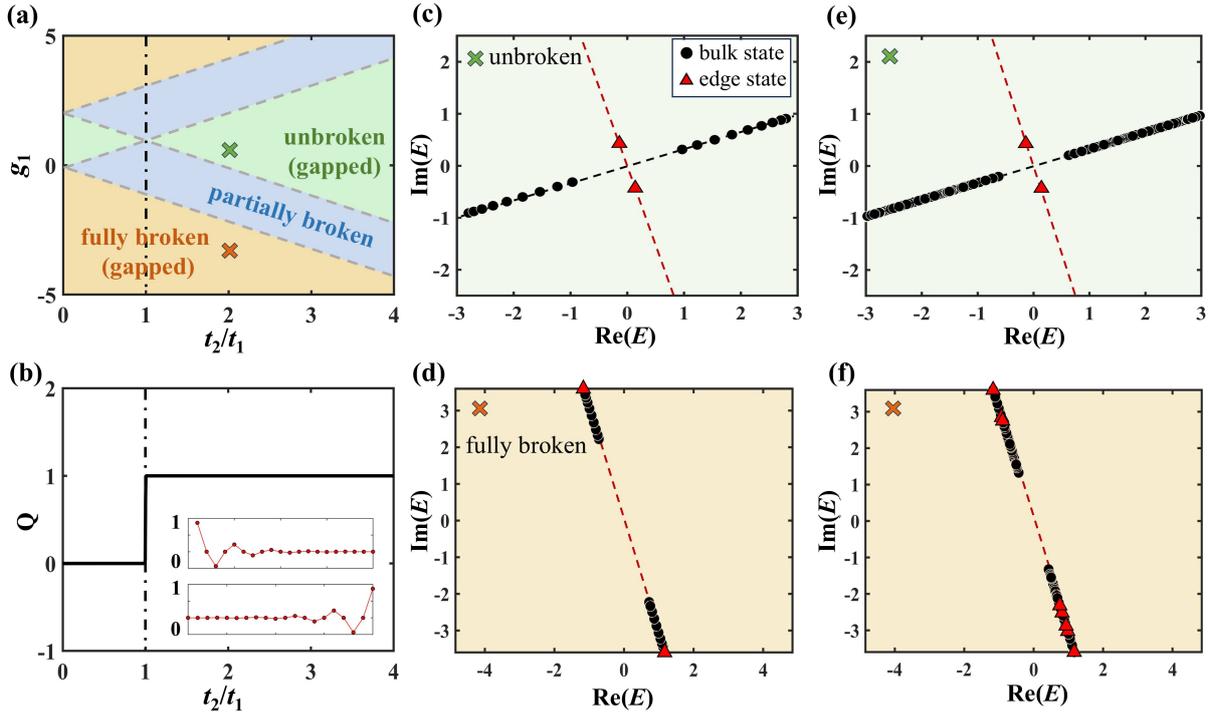

**Figure 4. Topological protection and robustness of the edge states.** (a) Phase diagram in the parameter space composed of $g_1$ and the coupling ratio $t_2/t_1$. The gray dashed lines represent the phase transition of anyonic-PT symmetry. Three colors represent the three phases of anyonic-PT symmetry. The vertical dotted–dashed line delineates the regions with and without edge states. (b) Calculated global Berry phase as a function of the coupling ratio for $g_1 = 0.5$. The insets show the profiles of two edge states when $Q = 1$, exhibiting chiral distributions. (c-

d) Eigenvalues of finite $H_0$ with $N = 20$ sites in (c) symmetry-unbroken regime with $g_1 = 0.5$ and (d) fully broken regime with $g_1 = -3$. Parameters correspond to the two colored crosses marked in panel (a) and $t_2 = 2t_1$. Red triangles denote the edge states and black circles denote the bulk states. (e-f) For systems in (c) and (d), random off-diagonal perturbations are applied for 100 realizations, preserving the PAH symmetry. In all figures, the angle between the black dashed line and the real axis is $\Gamma = 0.1\pi$, and $\omega_1 = 3$.

Although we only consider the case of $t_2 > t_1$ for analyzing the band structure of the infinite lattice here, the process of band gap closing and reopening exists as long as $t_2 \neq t_1$. The phase transition behavior under different parameter conditions can be obtained through the calculation of the Berry phase:[68, 69]

$$\varphi_B^\pm = \oint_C A^\pm = \oint_C i\langle u_\pm|\partial_k|\lambda_\pm\rangle, \tag{7}$$

where $A^\pm = i\langle u_\pm|\partial_k|\lambda_\pm\rangle$ is the Berry connection. Here $|\lambda_\pm\rangle$ and $\langle u_\pm|$ denote the right and left eigenstates of the Hamiltonian $H(k)$. Equation (7) shows the non-Hermitian induced geometric phase. Due to the gain-loss modulation in the system, the Berry phase of each band is a complex number. The phase transition of the anyonic-PT symmetry corresponds to the abrupt change in either the real or imaginary part of the Berry phase (see Supporting Information Figure S2). **Figure** 4a shows the phase diagram in the parameter space. The green and orange regions represent the unbroken and fully broken regimes of anyonic-PT symmetry, where the bulk bands are gapped. It can be observed that, for any coupling ratio (except $t_2 = t_1$), varying $g_1$ induces a gap closing and reopening.

By summing the Berry phases of the two bands, we can obtain the global Berry phase:[70]

$$Q = \varphi_B^+ + \varphi_B^-. \tag{8}$$

The value of $Q$ is quantized and can be used to distinguish the presence or absence of edge states. When $Q = 0$, the system has no edge states. When $Q = 1$, the system supports two edge states with chiral distributions, whose profiles are shown in the insets of Figure 4b. We find that $Q$ depends solely on the coupling ratio $t_2/t_1$. The topological phase transition, where $Q$ exhibits abrupt changes, is indicated by vertical dotted–dashed line located at $t_2 = t_1$ as shown in Figure 4a,b. We emphasize that two types of phase transitions are present. The first, indicated by the gray dashed line, separates different regimes of the anyonic-PT symmetry. The second, denoted by the vertical dotted–dashed line, distinguishes the presence or absence of edge states.

Here, we consider that the edge states are topologically protected only when $Q = 1$ and the system resides in anyonic-PT symmetry-unbroken regime. In conventional 1D PT or anti-PT-symmetric lattices, edge states are protected by pseudo-anti-Hermiticity or pseudo-Hermiticity symmetries.[67] However, these two symmetries are broken by the complex coupling term and the strict onsite potential constraints. We therefore introduce a generalized symmetry, namely, the pseudo-anyonic-Hermiticity (PAH) symmetry. It is such a new symmetry that protects the topological edge states in this system.

The PAH symmetry is defined as:

$$H_0(k) = -e^{2i\Gamma}\eta H_0^+(k)\eta^{-1}, \tag{9}$$

where $\eta = \eta^{-1} = \sigma_z$. $H_0^+(k)$ denotes the Hermitian conjugate of $H_0(k)$. Here, $H_0$ is not the original lattice Hamiltonian, but obtained by removing an overall energy shift from the original Hamiltonian H:

$$H_0 = H - \frac{\omega_1 + ig_1 + \omega_2 + ig_2}{2}. \tag{10}$$

The operation does not alter the topological properties but simplifies the analysis. Under the PAH symmetry, eigenvalues occur in pairs $(E, -E^* \exp(2i\Gamma))$. For a finite lattice, $\eta = \eta^{-1} = \sigma_z \otimes I_{N/2}$, where N is the total number of sites. As a result, the edge states exhibit a chiral distribution, i.e., they are eigenstates of η, residing one sublattice, as shown in the inset of Figure 4b. Due to boundary effects, the edge states satisfy $E = -E^* \exp(2i\Gamma)$, with their energies residing on the line at angle $\Gamma + \frac{\pi}{2}$ in the complex plane.

Figure 4c shows the eigenvalues of $H_0$ in anyonic-PT symmetry unbroken regime with $g_1 = 0.5$ and $t_2 = 2t_1$, corresponding to the parameter of green cross in Figure 4a. Bulk states (black circles) are constrained to lie along the black dashed line due to the anyonic-PT symmetry, while edge states (red triangles) are constrained by the PAH symmetry to lie along the red dashed line. In this case, edge states reside within the bulk line gap in the complex plane.[13]

Applying random off-diagonal perturbations that preserve PAH symmetry for 100 realizations (Figure 4e), the edge states remain in the gap and on the red dashed line, confirming that the edge states are protected by the PAH symmetry. In contrast, in the fully broken regime (Figure 4d), all bulk states collapse onto the red dashed line and the edge states merge into the bulk. The explicit form of the perturbation is given in Equation (S9) of the Supporting Information. Under perturbations (Figure 4f), although edge states remain phase constrained, they no longer reside in the gap and thus are not topologically protected.

Strictly speaking, PAH symmetry does not fix the edge state energies, but constrains their phases to be distributed along the red dashed line (see Figure 4c). Therefore, we can say that the PAH symmetry leads to a protected phase to $\Gamma + \frac{\pi}{2}$. Additional results for the diagonal perturbation are provided in the Supporting Information and show the same conclusion. Moreover, we find that regardless of the type of perturbation applied, as long as the absolute value of the intercell coupling exceeds that of the intracell coupling, edge states emerge and exhibit chiral profiles. Nevertheless, when the system is in anyonic-PT partially broken or fully broken regime, these edge states no longer reside in the gap and thus are not topologically protected. They are topologically protected only when $Q = 1$ and the system is in the anyonic-PT unbroken regime.

## 5. Conclusion

In summary, we have investigated the role of anyonic-PT symmetry in non-Hermitian waveguide systems, with particular emphasis on its impact on phase transitions and topological properties. By analyzing both a minimal two-site model and an extended one-dimensional lattice, we uncover distinctive spectral and dynamical features that set anyonic-PT symmetry apart from conventional PT and anti-PT-symmetric systems. In the symmetry-unbroken regime, the bulk eigenvalues are constrained to lie along a tilted line in the complex plane, while the eigenvalues of edge states are pinned to a perpendicular line, reflecting the coexistence of multiple symmetry constraints.

Our study of the extended lattice further reveals a direct and nontrivial connection between anyonic-PT symmetry and the emergence of topological boundary states. We show that these boundary states arise from a gap-closing and reopening process in the complex band structure and are robust only within the anyonic-PT symmetry-unbroken regime. Their protection is guaranteed by a generalized symmetry termed the pseudo-anyonic-Hermiticity symmetry, which constrains the phase of the edge state energies rather than fixing their absolute values. This mechanism provides a deeper understanding of topological protection in non-Hermitian systems and highlights the role of eigenvalue arguments as physically meaningful and controllable quantities.

The rich physics enabled by anyonic-PT symmetry suggests promising opportunities for applications in optical signal manipulation and information processing, where the eigenvalue phase may serve as an additional degree of freedom beyond amplitude and frequency. At the same time, many aspects of anyonic-PT symmetry remain unexplored. In particular, the enhanced sensitivity associated with exceptional points, as well as the possibility of dynamically tuning and controlling topological localized states through anyonic-PT parameters or optical nonlinearities, merit further investigation. Our work thus establishes a foundation for exploring generalized non-Hermitian symmetries, phase-engineered topology, and their applications in photonic systems.

**Acknowledgements**

This work was supported by the National Key R&D Program of China (2022YFA1404800); National Natural Science Foundation of China (12134006, W2541003, 12374309, 12274242 and 124B2078); Natural Science Foundation of Tianjin (21JCJQJC00050); the 111 Project (B23045) in China; Institute of Electronic Structure and Laser, FORTH and the University of Crete by the European Research Council (ERC-Consolidator) under the grant agreement No. 101045135 (Beyond_Anderson).

**Conflict of Interest**

The authors declare no conflict of interest.

**Data Availability Statement**

The data that support the findings of this study are available from the corresponding author upon reasonable request.

**Phase Transitions and Topological Protection in Anyonic-PT-Symmetric Lattices**

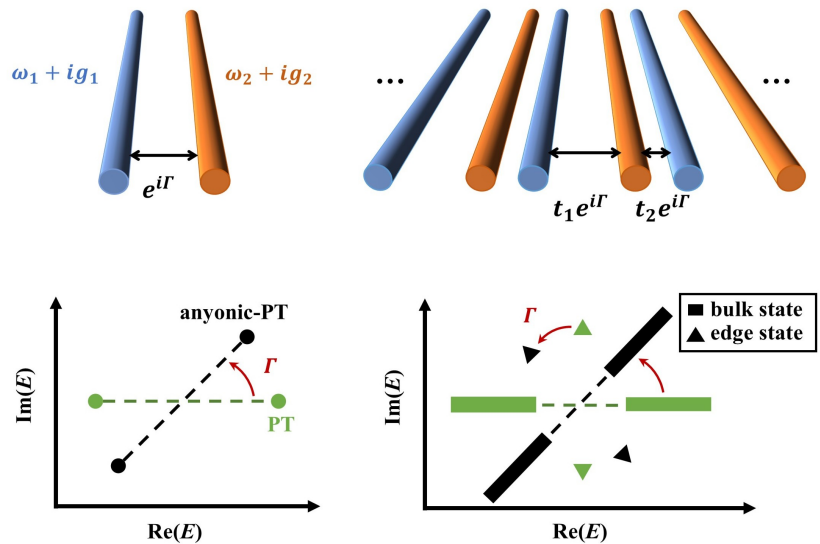

Anyonic-PT symmetry extends PT physics through a tunable phase $\Gamma$ that rotates the complex spectrum. Using a two-site dimer and an extended 1D lattice, we investigate the phase transition and further characterize it by the Berry phase. In the symmetry-unbroken regime, we explore the dynamics and show that edge states are protected by pseudo-anyonic-Hermiticity symmetry.